Title
- Fast and label-free 3D virtual H&E histology via active modulation-assisted dynamic full-field OCT

Authors

Zichen Yin[1,2],†, Bin He[1,2],†, Yuzhe Ying[3], Shuwei Zhang[4], Panqi Yang[1,2], Zhengyu Chen[1,2], Zhangwei Hu[1,2], Yejiong Shi[1,2], Ruizhi Xue[1,2], Chengming Wang[1], Shu Wang[4]*, Guihuai Wang[3]* & Ping Xue[1,2]*

Affiliations
[1] *State Key Laboratory of Low-dimensional Quantum Physics and Department of Physics, Tsinghua University, Beijing, 100084, China.*
[2] *Frontier Science Center for Quantum Information, Beijing, China.*
[3] *Department of Neurosurgery, Beijing Tsinghua Changgung Hospital, School of Clinical Medicine and Institute of Precision Medicine, Tsinghua University, Beijing, 102218, China.*
[4] *Breast Center, Peking University People's Hospital, Beijing 100044, China.*
†These authors contributed equally to this work.
* *Corresponding author: Shu Wang email: shuwang@pkuph.edu.cn;*
 *Guihuai Wang email: youngneurosurgeon@163.com;*
*Ping Xue email: xuep@tsinghua.edu.cn.*

Abstract

Pathological features are the gold standard for tumor diagnosis, guiding treatment and prognosis. However, standard histopathological process is labor-intensive and time-consuming, while frozen sections have lower accuracy. Dynamic full-field optical coherence tomography (D-FFOCT) offers rapid histologic information by measuring the subcellular dynamics of fresh, unprocessed tissues. However, D-FFOCT images suffer from abrupt shifts in hue and brightness, which is confusing for pathologists and diminish their interpretability and reliability. Here, we present active phase modulation-assisted D-FFOCT (APMD-FFOCT) to improve the imaging stability and enhance the contrast of static tissues. This enables us to further employ an unsupervised deep learning to convert APMD-FFOCT images into virtual hematoxylin and eosin (H&E) stained images for the first time. Three-dimensional (3D) virtual H&E-stained images have been obtained at a scanning rate of 1 frame per second, as demonstrated in cancer diagnosis for human central nervous system and breast. The results prove that this new method will play a unique and important role in intraoperative histology.

Teaser
3D virtual H&E staining from subcellular dynamics insights through unsupervised deep learning.

MAIN TEXT

Introduction

Pathological features serve as the gold standard for tumor grading, forming the cornerstone of treatment decisions (*1*). During surgical interventions, the choice of surgical strategy is often guided by the findings from pathological examinations (*2*). This involves striking a delicate balance between aggressive and conservative tumor resection strategies based on the malignancy level of the tumor, especially in critical body parts like central nervous system (CNS). An excessively aggressive resection strategy can compromise the patient's normal functions, causing additional suffering. Conversely, a too-conservative strategy risks incomplete tumor excision and high tumor recurrence rates (*3*). The surgical goal is to remove as much of the tumor tissues as possible while preserving functional tissues. Hence, the ability to obtain high-quality pathological images in a timely and convenient manner during surgery is crucial for making intraoperative decisions and ensuring clean surgical margins (*3*). However, the standard histopathological biopsy process is laborious and time-consuming, involving formalin fixation and paraffin embedding (FFPE), followed by thin sectioning, staining, and mounting on glass slides, often extending over several days (*4*). Frozen sectioning provides a rapid intraoperative diagnosis method, but this technique can introduce severe artefacts, such as freezing artefacts, poor quality sectioning, swollen cell morphologies and poor staining. These effects are especially pronounced in tissues that do not freeze well, like those in breast and brain, resulting in reduced diagnosis accuracy (*5, 6*).

In order to overcome these drawbacks of standard and frozen histology and provide rapid intraoperative diagnosis, high-resolution and high-contrast "optical biopsy" techniques that reduce or obviate the need for biological tissue processing are currently being explored. Among these novel imaging methods, one category utilizes point-by-point scanning for image acquisition, such as stimulated Raman scattering (*7, 8*), confocal microscopy (*9, 10*), multiphoton microscopy (*11-14*), second-harmonic imaging microscopy (*15*) and photoacoustic microscopy (*16, 17*). Most of these imaging modalities utilize a highly focused laser beam to induce linear and nonlinear light-matter interactions, allowing for high-resolution, label-free imaging. However, the requirement to scan across three dimensions and the time needed to linger on each pixel for data acquisition means that assembling a three-dimensional image is a very time-consuming process (*18*). In addition, these approaches can lead to phototoxicity and photodamage. Another category including wide-field structured illumination microscopy (SIM) and light-sheet microscopy like open-top light-sheet (OTLS) microscopy and MediSCAPE system offers significant advantages in imaging speed as they can capture an entire plane in each acquisition (*18-20*). However, these techniques involve complicated geometric configurations, which add to the complexity and cost of the system. In addition, they typically rely on exogenous fluorochromes to enhance their signal-to-noise ratio (SNR) and enrich their contrast, which increases the complexity of tissue processing and potentially affects the subsequent tissue reutilization.

In recent years, dynamic full-field optical coherence tomography (D-FFOCT) has been developed, utilizing subcellular dynamics as an intrinsic contrast source to enable high contrast imaging (*21, 22*). This technique measures variations in the optical path length of backscattered light by capturing the interference signal with a camera, thus sensitively responding to the nanoscale movements of subcellular particles across the entire field of view. By extracting statistical features of the interference signal, optical histology can be achieved without any tissue processing. Offering improved image contrast over traditional FFOCT (*23, 24*), D-FFOCT has demonstrated high accuracy in the diagnosis of breast cancer and nodal metastasis (*25*). However, axial displacements of the sample caused by environmental vibration may often mix with intracellular motions, resulting in a reduction of the SNR in D-FFOCT images (*26*). And because of the random nature of environmental noises, consecutive D-FFOCT images of the adjacent tissues often exhibit random shifts in hue and brightness (*25*). Since the hue and brightness of D-FFOCT images reflect metabolic indexes of tissues, this instability will confuse pathologists and reduce the interpretability of the images. In addition, pathologists are typically trained to analyze standard H&E histology for diagnosis, hence unfamiliar to D-FFOCT colored images. Therefore, D-FFOCT may raise the learning threshold for pathologists and impede fast decision making.

To overcome these barriers, pseudo-H&E images have been developed based on linear (*27*) and nonlinear mapping (*28*) for other optical imaging techniques. But these methods usually require fluorescent dyes and result in unnatural-looking images. Recently, deep learning (*29, 30*) has been applied to generate virtual H&E histology from various imaging modalities (*31-42*). However, the performance of virtual H&E histology significantly depends on the quality and characteristics of the original images used for training. For example, imaging techniques, such as autofluorescence imaging (*31, 37*), quantitative phase images (QPI) (*32*), reflectance confocal microscopy (RCM) (*33*), quantitative oblique back illumination microscopy (qOBM) (*38*), stimulated Raman scattering (SRS) microscopy (*39*) and point-scanning OCT (*40*) fail to provide sufficient nuclear contrast, especially lacking intranuclear details. Consequently, errors in nucleus generation or omissions can occur after deep learning transformations. Microscopy with UV surface excitation (MUSE) (*41, 42*) and UV photoacoustic microscopy (PAM) (*16, 17, 34*) can provide nuclear contrast within tissues. However, MUSE requires dye staining of the sample and lacks depth-resolving capability, while PAM cannot resolve individual cell nuclei within densely packed cells due to limited axial resolution. In fact, D-FFOCT could provide high-contrast and high-resolution nuclear structures without labeling, but its instability in hue and brightness as mentioned above prohibits the correct generation of virtual H&E histology.

In this study, we firstly utilize a novel technique, as called APMD-FFOCT that adopts active phase modulation to eliminate the influence of the random environmental vibration. We demonstrate that this method significantly enhances the stability of dynamic images and achieves continuity and consistency for image stitching. Meanwhile, this method also offers good contrast to static tissues, such as tumor-associated collagen fibers and calcified tissues, while maintaining high sensitivity to dynamic tissues like tumor cells. Therefore, the APMD-FFOCT images provide much more solid foundation than conventional ones for virtual H&E-staining. We demonstrate, for the first time, that APMD-FFOCT images can be converted into virtual H&E-stained images by unsupervised deep learning, achieving three-dimensional virtual H&E-stained images at a scanning rate of 1 frame per second. Furthermore, we also show that this novel technique has successfully applied in cancer diagnosis for human central nervous system and breast. These results demonstrate that our method has the potential for scanning large, thick tumor tissues intraoperatively, making it an ideal tool in intraoperative histology.

## Results
### APMD-FFOCT system and imaging principle.

As shown in Fig. 1A, the APMD-FFOCT configuration utilizes a Linnik interferometer with twin 20x water immersion objectives and a low coherence light source. The light returning from the sample layer interferes with the light reflected back from the reference mirror and is subsequently projected onto the camera. More system specifics are given in the Methods section. APMD-FFOCT detects the subcellular movements by capturing changes in the optical length of back-scattered light from actively metabolizing, freshly excised tissues. The fluctuation of light intensity on each pixel of the camera is recorded as a time-domain intensity traces, as shown in Fig. 1B. These traces then undergo Fourier transformation to generate their corresponding frequency spectra. We then employed the Hue-Saturation-Value (HSV) color space to visualize the dynamic characteristics of the metabolism activities. As shown in Fig. 1C, hue is determined by the spectral centroid of the frequency spectrum, thus reflecting the fluctuation speed. The brightness of the image is determined by the integration of the frequency spectrum without the zero-frequency portion, which links to the fluctuation amplitude. As shown in Fig. 1A, a sinusoidal signal can be added to the PZT under reference mirror to generate active phase modulation. With a chosen frequency of 25 Hz and amplitude of 40 nm, this active phase modulation forms a sharp and stable intensity peak in frequency spectrum, as shown in Fig. 1C.

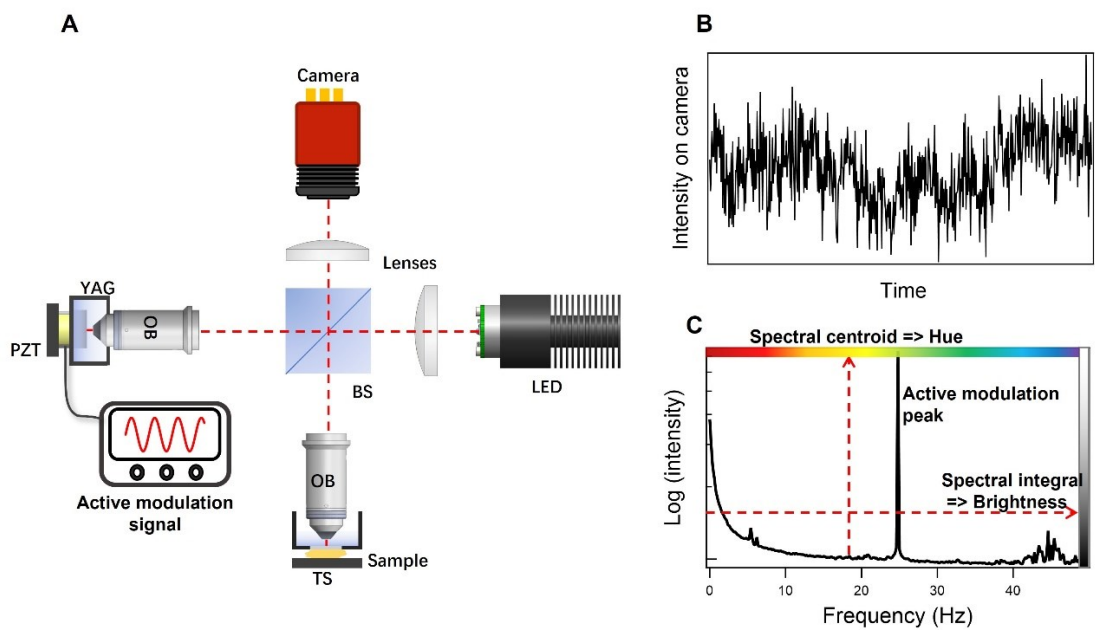

**Fig. 1. APMD-FFOCT system designs and image generation.** (**A**) Schematic of active phase modulation-assisted D-FFOCT. BS, Beam splitter; OB, Objectives; PZT, piezoelectric translation; TS, translation stage. YAG, Yttrium Aluminum Garnet. The signal generator provides an active modulation signal to the PZT in reference arm. (**B**) An intensity trace that records the temporal fluctuations of back-scattered light from a pixel of the camera. (**C**) Frequency spectrum obtained by Fourier transform of the intensity trace.

**Instability in D-FFOCT images and the implementation of active phase modulation correction**

Fig. 2A displays a portion of the stitched D-FFOCT image of human psammomatous meningioma, highlighting abrupt changes in both hue and brightness of consecutive D-FFOCT images. To elucidate this phenomenon, the averaged frequency spectra were extracted from the selected areas, as depicted in Fig. 2B. The spectrum integrals and centroids of ROI 1, ROI 2, and ROI 3 in Fig. 2A were 64 and 25.6 Hz, 54 and 26.4 Hz, and 67 and 24.8 Hz, respectively. The instability of these dynamic indexes is thus mainly due to the random distribution of intensity peaks in the high frequency range of the spectra in Fig. 2B. Moreover, Fig. 2C shows the stitched D-FFOCT image with both collagen fibers and tumor cells. In the two adjacent images, while the hue and brightness for tumor cells remain consistent, a sudden change is observed in the collagen fibers. Fig. 2D exhibits the averaged frequency spectra of tumor cells and collagen fibers in solid boxes of Fig. 2C. The results reveal that the spectral intensity peaks for both collagen fibers and tumor cells occur at identical frequencies but with significant variations in intensity, as marked by the arrows.

This implies that the intensity peaks stem from the bulk vibration of the sample rather than intrinsic subcellular motions. Furthermore, the variation in peak intensities suggests that collagen fibers have much higher reflectivity than tumor cells, making them more susceptible to ambient vibrations. Therefore, the random distribution of environmental vibrations results in the instability of hue and intensity in the highly reflective sections of tissues.

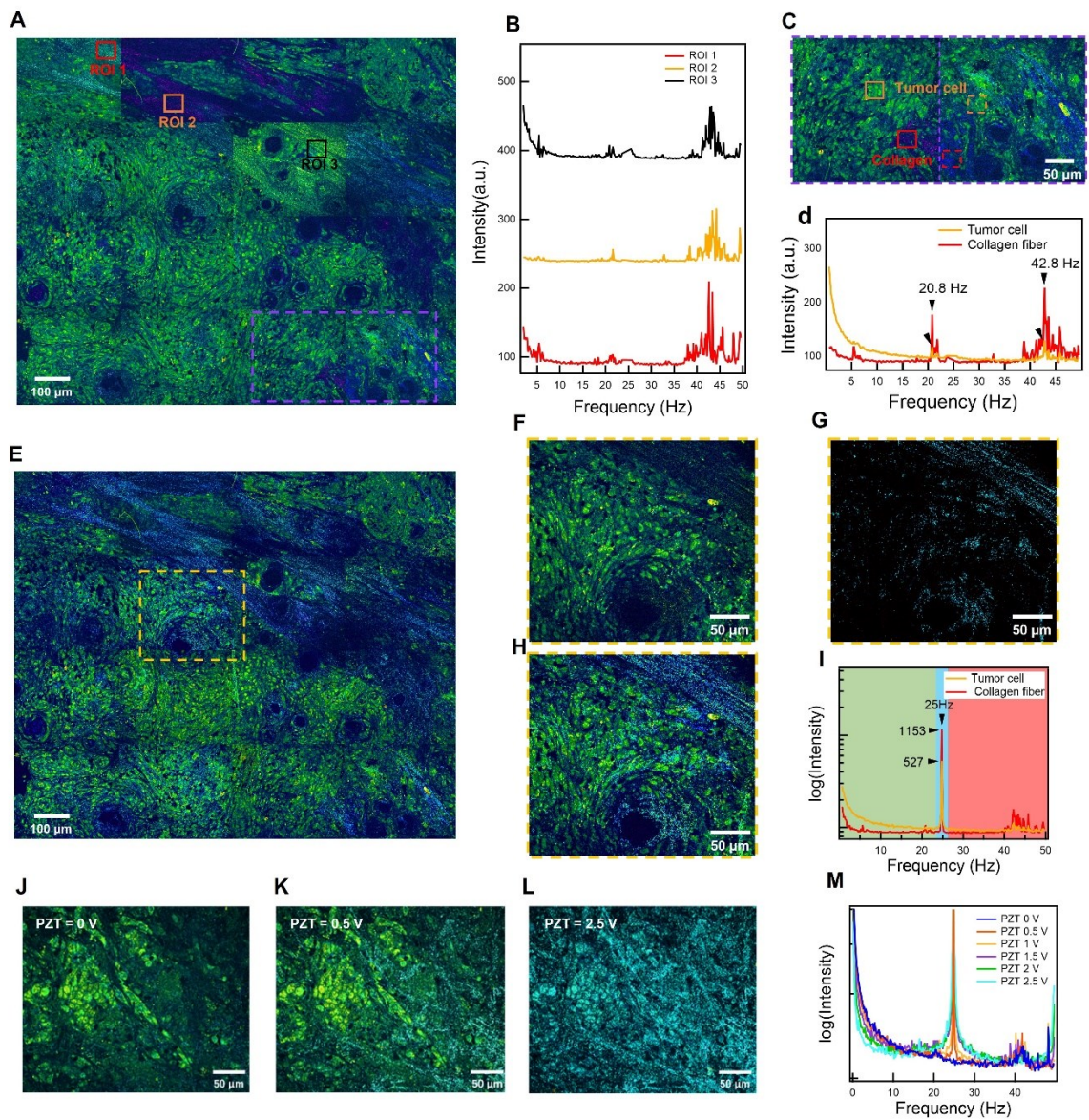

**Fig. 2. Instability in D-FFOCT images and the implementation of active phase modulation correction.** (**A**) Stitched D-FFOCT image of psammomatous meningioma with collagen fibers. (**B**) Averaged frequency spectra correspond to the selected regions in **A**. (**C**) Stitched D-FFOCT image with both collagen fibers and tumor cells (enlargement of purple dotted box in **A**). (**D**) Averaged frequency spectra correspond to the selected solid boxes of collagen fibers and tumor cells in **C**. (**E**) Stitched APMD-FFOCT image of the same area in **A**. (**F**) Image generated by low-frequency part (0-24 Hz, green area of **I**), (**G**) Image generated by active phase modulation part (24-26 Hz, blue area of **I**). (**H**) the combination of **F** and **G**. (**I**) Averaged frequency spectra correspond to the collagen fibers and tumor cells in **H**. (**J-L**) APMD-FFOCT images with PZT modulation voltage amplitude of 0 V (**J**), 0.5 V (**K**) and 2.5V (**L**). (**M**) Frequency spectra of tumor cells areas in **J-L** with increasing PZT modulation voltage.

A straightforward method to mitigate these abrupt changes in hue and brightness is to exclude the high-frequency portion of the frequency spectrum, as marked by the red segment in Fig. 2I. The green segment, denoting the metabolic frequency region, was employed to generate Fig. 2F. In this figure, only the tumor cells are visible. However, collagen fibers, which are not evident in this representation, are vital as they are one of the tumor features and crucial for providing accurate pathological information. Unfortunately, the intrinsic motion of the collagen fibers is too faint to be effectively detected. Therefore, to enhance the contrast of collagen fibers, we introduce a 25 Hz sinusoidal signal to the PZT to actively modulate the reference mirror while capturing intracellular motions. This active phase modulation results in a sharp but stable intensity peaks in frequency spectra at 25 Hz, as shown in the blue area of Fig. 2I. The integral of the blue area of the frequency spectrum is used to generate Fig. 2G, where collagen fibers are effectively preserved due to its high reflectivity. Fig. 2H presents the combination of Fig. 2F and Fig. 2G. In this figure, the hue and brightness of the collagen fibers can be flexibly and independently adjusted for contrast optimization and remain stable across different images. It should be noted that the collagen fiber portion is hardly visible in Fig. 2F due to its low dynamics in traditional D-FFOCT but quite clear in Fig. 2H due to active phase modulation in APMD-FFOCT. Furthermore, stitched images from same tissue region without and with active phase modulation, as shown in Fig. 2A and Fig. 2E, obviously prove that the stability of hue and brightness of APMD-FFOCT image is significantly improved with enhanced contrast of collagen fibers.

In order to optimize image quality, we conducted a comparative study to assess the impact of PZT modulation voltage when imaging the same tissue region. Fig. 2J-L represent APMD-FFOCT images with PZT modulation voltages of 0 V (no modulation), 0.5 V (active phase modulation voltage) and 2.5 V (traditional FF-OCT phase modulation voltage). The corresponding frequency spectra with varying PZT modulation voltage are summarized in Fig. 2M. Comparing Fig. 2J to Fig. 2K, it is evident that the latter offers clearer collagen fiber structures. Corresponding frequency spectrum curves of 0 V and 0.5 V modulation voltages show that their low-frequency curves are nearly identical (Relative difference of spectrum integrals: ~1％), indicating that a 0.5 V modulation voltage has a negligible impact on the detection of cellular dynamics. The frequency spectrum curve of 0.5 V modulation voltage exhibits a sharp intensity peak near 25 Hz. Upon further increasing the modulation voltage to 2.5 V, as shown in Fig. 2L, it is evident that the image contrast is reduced, and the signals from active phase modulation gradually overwhelm the signals from cellular dynamics. Corresponding frequency spectrum curves from 1 V to 2.5 V modulation voltage also reveal an increasing suppression of cellular dynamics in the low-frequency range. Therefore, we choose PZT modulation voltage of 0.5 V for our APMD-FFOCT.

**Comparison of APMD-FFOCT and H&E-stained images in CNS tumors**

CNS tumors encompass various pathological entities, each displaying unique histological features. Fig. 3A presents APMD-FFOCT images of psammomatous meningioma. Its histological signature includes numerous psammoma bodies—concentrically calcified structures within the tumor tissue. These bodies, being relatively static, are either not visible or appear unstable in traditional D-FFOCT images. However, active phase modulation makes these structures clear and distinct in APMD-FFOCT images, ensuring high consistency and continuity in the stitched images. Fig. 3D displays corresponding H&E-stained images from the same sample, showing similar psammoma body structures. Yet, in H&E-stained images, most psammoma bodies appear fragmented and incomplete. Fig. 3J zooms into the area indicated by the blue box in Fig. 3D, highlighting the out-of-focus blur of remaining psammoma bodies. These issues may arise from the inherent hardness of the psammoma bodies, which, when subjected to microtome slicing, can lead to fragmentation and detachment from the tissue section. Additionally, this process can also result in uneven of the remaining psammoma bodies and lead to the out-of-focus blur. These problems are completely avoided in slide-free APMD-FFOCT imaging. Moreover, nuclei from round to oval in shape in APMD-FFOCT image are clearly discernible in Fig. 3G, consistent with H&E-stained images in Fig. 3J. APMD-FFOCT and H&E-stained images of ependymoma are shown in Fig. 3B and 3E, respectively. The collagen fibers in both images are distinctly visible, while highlighted in blue in APMD-FFOCT image, weaving through the regions of tumor cells. The tumor nuclei, from round to slightly oval with moderate density, are visible in Fig. 3H and 3K. Fig. 3C and 3F reveal APMD-FFOCT and H&E-stained images of schwannoma, where Antoni A areas can be identified by densely packed spindle-shaped cells with elongated nuclei (Fig. 3I and 3L), which are marked features of schwannomas(*43*). In conclusion, despite from different contrast, APMD-FFOCT has demonstrated its ability to reveal tissue and cellular details comparable to those seen in H&E-stained images.

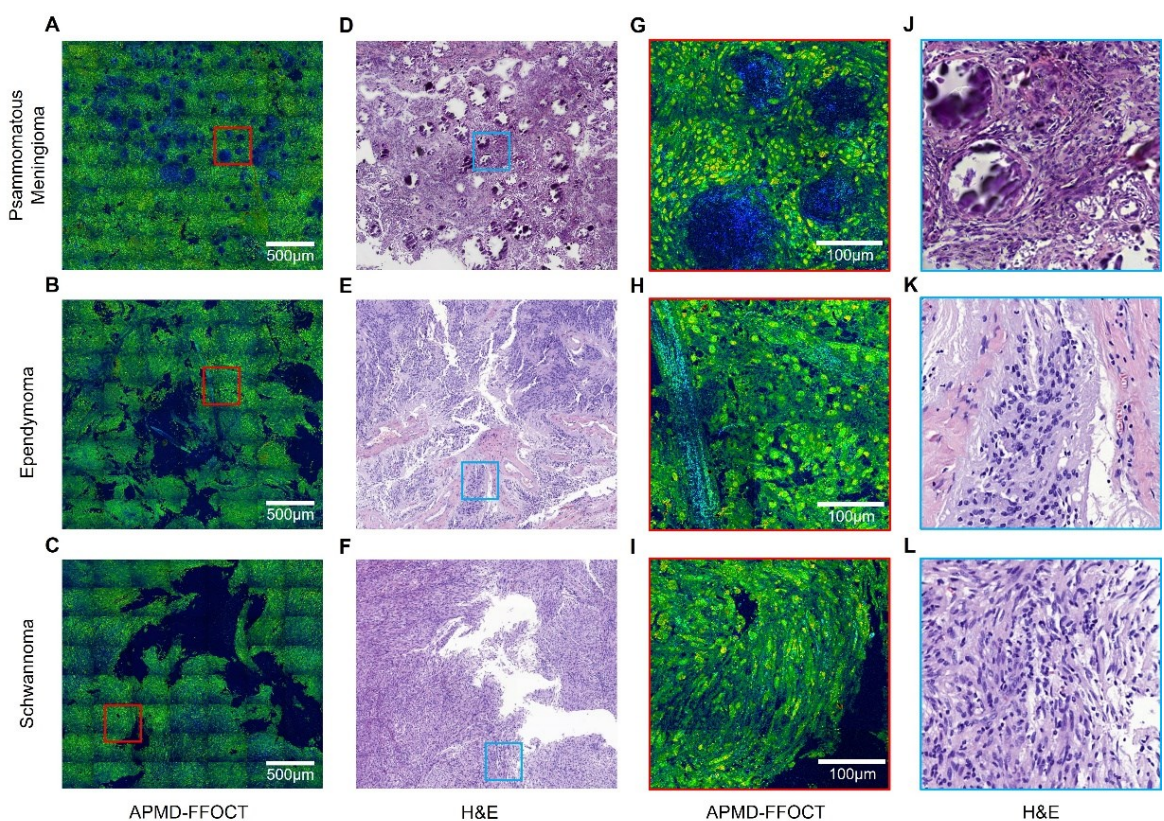

**Fig. 3. Comparison of APMD-FFOCT and H&E staining images in CNS tumors.** (**A-C**) Stitched APMD-FFOCT images of psammomatous meningioma (**A**), ependymoma (**B**) and schwannoma (**C**). (**D-F**) H&E-stained images of the same sample correspond to **A-C**. (**G-I**) Enlargement of the red boxes in **A-C**. (**J-L**) Enlargement of the blue boxes in **D-F**.

**Virtual staining of APMD-FFOCT images of diffuse midline glioma**

Although APMD-FFOCT offers rapid access to high-contrast pathological images, it concurrently elevates the learning curve for pathologists and hampers rapid decision-making. This challenge arises because pathologists are traditionally trained to diagnose through standard histological images. After establishing the capability to capture stable dynamic images with histological details comparable to H&E-stained images, we applied a CycleGAN-based deep learning approach to perform virtual H&E staining on APMD-FFOCT images. More detailed procedures are described in Methods. Fig. 4A shows the APMD-FFOCT image of diffuse midline glioma (DMG), where tumor cells are densely arranged, with significant nuclear pleomorphism. Fig. 4B displays virtual H&E-stained images derived from Fig. 4A, in which the histological features are accurately converted into a representation resembling traditional H&E staining (Fig. 4C). Fig. 4D-F show enlargements of corresponding boxes in Fig. 4A-C. The morphology of the tumor cells is accurately translated, as exemplified by the cells to which the red arrows are pointing. Meanwhile, the tumor cells in the virtual H&E-stained images demonstrate excellent concordance with the style and appearance of traditional H&E images (Fig. 4F). Moreover, the dark corners in stitched APMD-FFOCT images (Fig. 4D), caused by uneven illumination and interference, are largely corrected (Fig. 4E). Each APMD-FFOCT image in Fig. 4 was calculated from 500 raw images obtained in 5 seconds. The acquisition time was set to 5 seconds, as extending it further was deemed unlikely to notably enhance the image SNR and quality. However, moderately reducing the acquisition time and the number of raw images can still produce APMD-FFOCT images with sufficient quality for further deep learning transformations. Fig. 5A-C present APMD-FFOCT images generated from 100, 300, and 500 raw images, captured over durations of 1, 3, and 5 seconds, respectively. As the acquisition time and number of raw images are reduced, there is an increase in speckle noise and a decrease in the SNR. Fig. 5D-F display virtual H&E-stained images derived from Fig. 5A-C. Through the conversion process, speckle noise is effectively removed, leading to virtual H&E-stained images in Fig. 5D-F that show uniform quality without obvious differences.

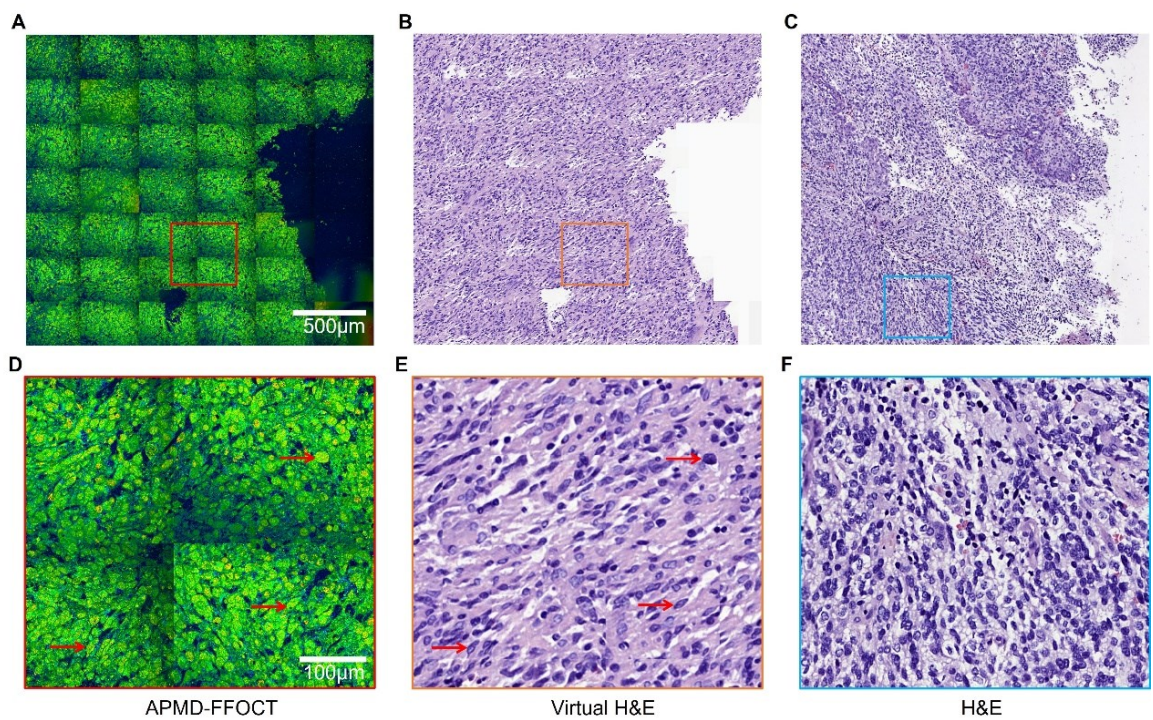

**Fig. 4. Conversion of APMD-FFOCT images to virtual H&E-stained images of DMG.** (**A-C**) Stitched APMD-FFOCT images, virtual H&E-stained images and standard H&E-stained images of DMG. (**D-F**) Enlargement of the selected regions in **A-C**.

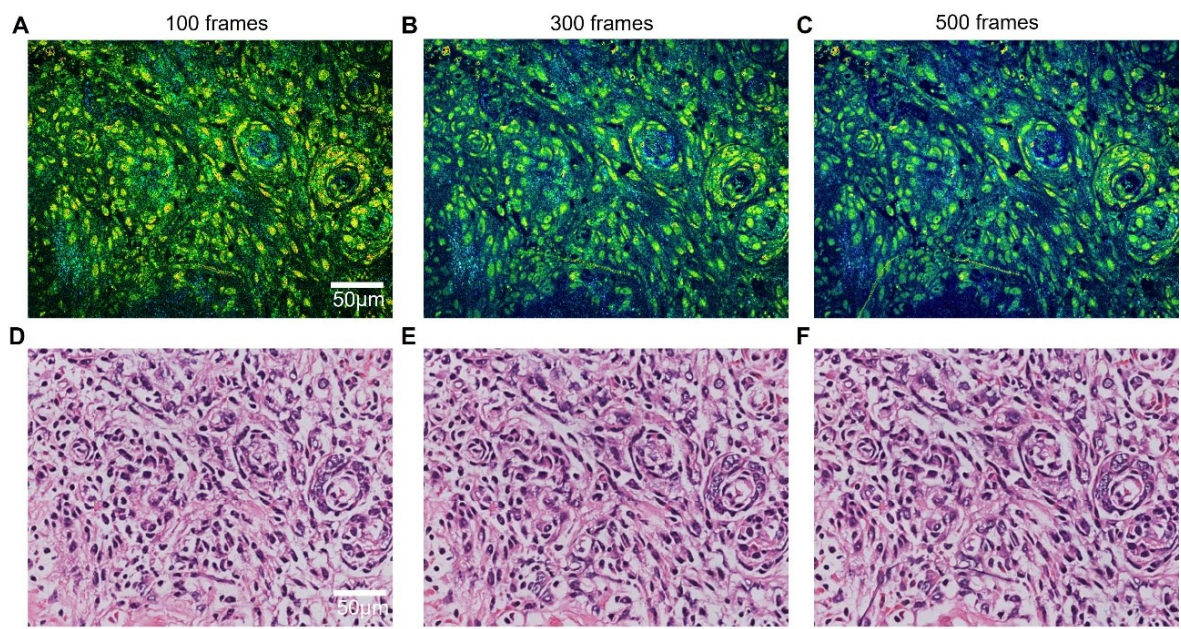

**Fig.5 Virtual H&E staining converted from APMD-FFOCT images generated by different original images.** (**A-C**) APMD-FFOCT images generated by 100, 300 and 500 raw images. (**D-F**) virtual H&E-stained images converted from **A-C**.

## APMD-FFOCT to virtual H&E-stained images conversion of invasive ductal carcinoma (IDC)

Having confirmed the capability to convert APMD-FFOCT images of CNS tumors into virtual H&E-stained images, we now proceed to apply this approach to breast tumors. Breast cancer is not only one of the most common cancers affecting women worldwide but also a leading cause of cancer-related deaths in females (*44*). IDC is the most common type of breast cancer, whose diagnosis and grading heavily depend on H&E staining. The Nottingham histological grading system for breast cancer emphasizes nuclear size and pleomorphism as critical factors for tumor grading (*45*). Furthermore, the collagen fibers serve as a prognostic indicator for survival in breast carcinoma (*46*), offering a more comprehensive understanding in diagnosis. We show that APMD-FFOCT images of IDC samples in grade III, grade II and grade I can be accurately translated into virtual H&E-stained images, providing rich pathological features like morphology of nucleus and distribution of collagen fibers.

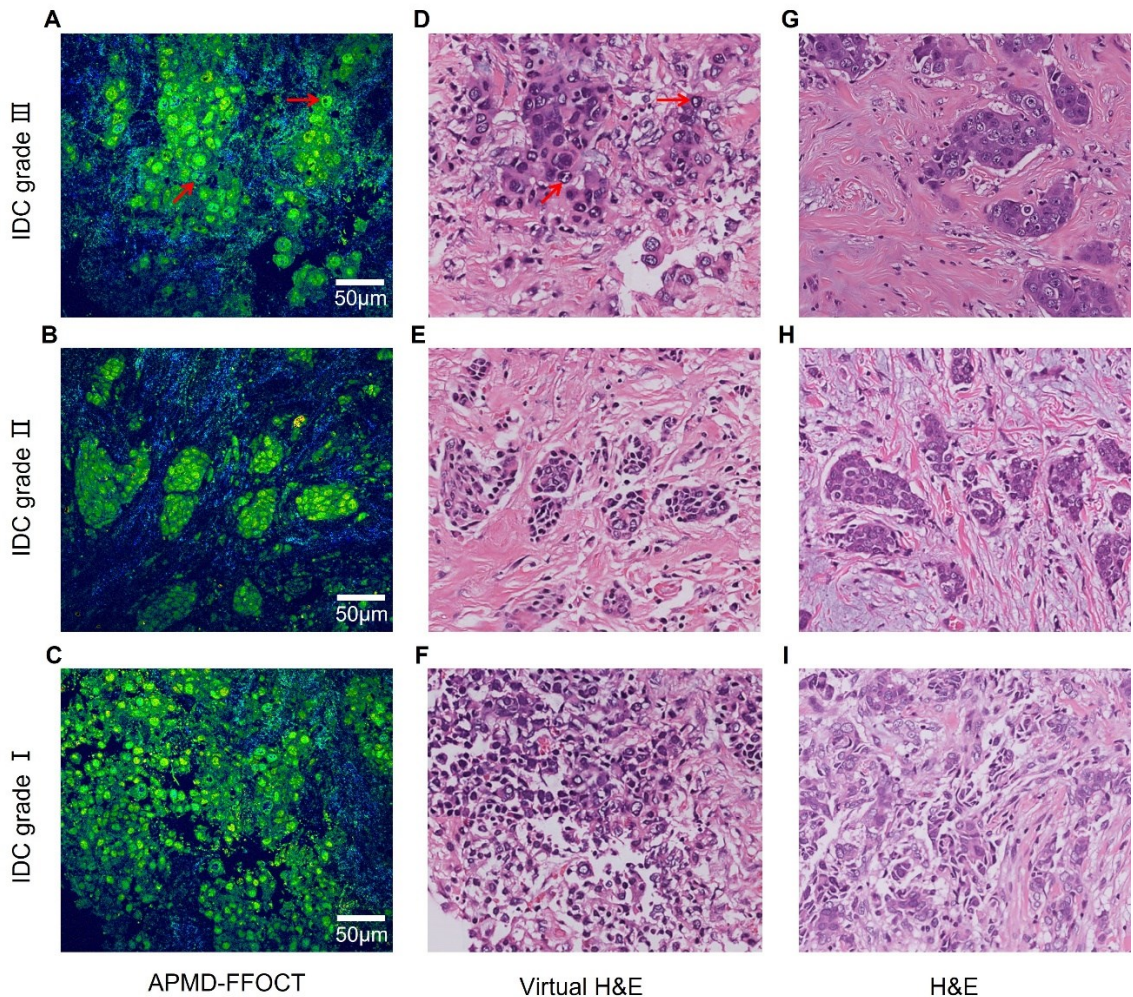

**Fig. 6 APMD-FFOCT to virtual H&E-stained images conversion of IDC.** (**A-C**) APMD-FFOCT images from IDC samples of grade III, grade II and grade I. (**D-F**) Virtual H&E-stained images converted from **A-C**. (**G-I**) Standard H&E-stained images of the same samples correspond to **A-C**.

As illustrated in Fig. 6A, APMD-FFOCT image of grade III IDC shows a large nuclear area with more pronounced nuclear pleomorphism. Fig. 6D displays the virtual H&E-stained images derived from Fig. 6A, wherein the size and morphology of the nuclei are faithfully mapped. Especially, the subnuclear details can also be resolved and mapped as pointed by the arrows. The second row of Fig. 6 depicts a grade II IDC sample, showing densely packed tumor cells that are embedded within the collagen fibers. Fig. 6B captures the intricate details of collagen fibers, rendered in blue. The distribution and orientation of collagen fibers are faithfully mapped in virtual H&E-stained image in Fig. 6E. The virtual H&E staining and standard H&E staining show high consistency (Fig. 6E and Fig. 6H). Furthermore, Fig. 6C shows APMD-FFOCT image of grade I IDC sample, also proving the feasibility of our method. The tumor cell nuclei are bright and easily distinguishable, surrounded by a comparatively dimmer cytoplasm, both exhibiting yellow-green hues. Correspondingly, the virtual H&E-stained image in Fig. 6F highlights the nuclei of tumor cells in shades of purple and the cytoplasm in lighter. In fact, the accuracy of the transformations for each of the above parts are obviously validated in the standard H&E-stained images shown in Fig. 6G-I.

**3D Virtual H&E Staining of APMD-FFOCT**

Due to low coherence gating, APMD-FFOCT enables high axial resolution and label-free tomographic imaging without the need of slicing. We can acquire three-dimensional tomographic images without any damage to the sample. Fig. 7A displays a volumetric 3D APMD-FFOCT tomography of a grade III IDC sample, which is composed of 50 two-dimensional slices, each separated by an axial distance of 0.3 micrometers with a volume of 350 μm × 500 μm × 15 μm. This high-resolution stack not only allows for an in-depth study of tissue morphology but also enables detailed analysis of nuclear structures of any cross-section at arbitrary angle. For example, the cross section in Fig. 7A shows clear longitudinal profile with rich intranuclear structures, which provides a unique perspective for doctors to analyze pathological details. Fig. 7B shows APMD-FFOCT image slices and the corresponding virtual H&E image slices at various depths. As APMD-FFOCT enables high contrast and high SNR imaging at various tissue depths, all APMD-FFOCT images are accurately transformed into virtual H&E images. The details inside nuclei shown as dark, indicated with arrows can be successfully translated into the white areas of nuclei in virtual H&E-stained images. Therefore, the first 3D virtual H&E Staining of APMD-FFOCT is successfully achieved, which will play a new unique and important role in intraoperative histology.

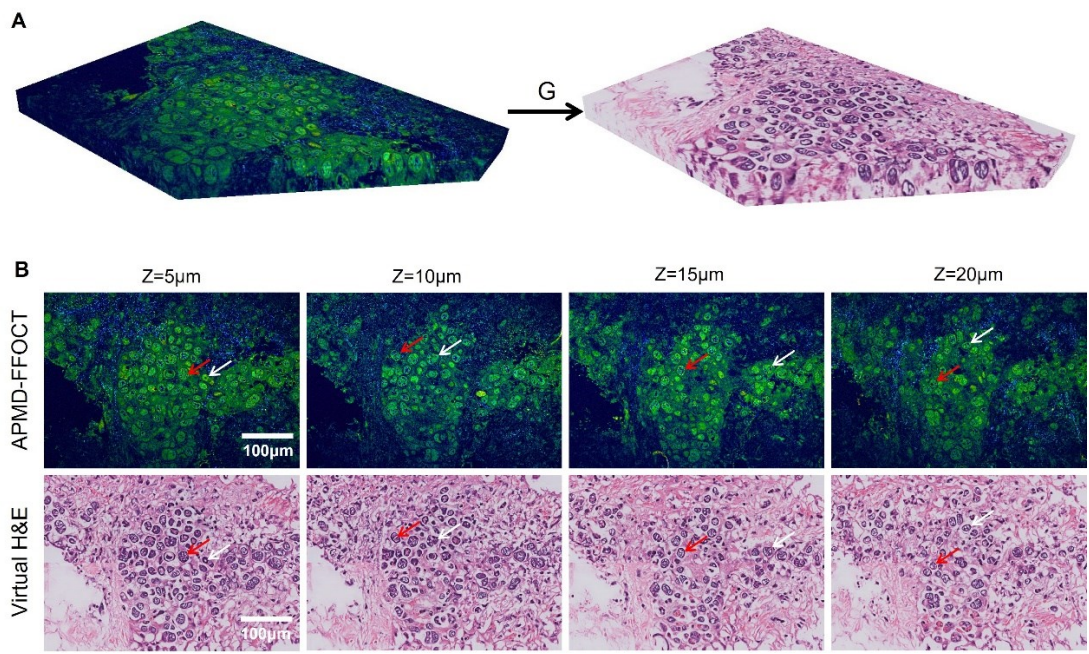

**Fig. 7. Virtual H&E Staining of 3D tomographic images.** (**A**) 3D APMD-FFOCT volume to 3D virtual H&E-stained volume conversion of IDC sample. (**B**) Comparation of APMD-FFOCT images and corresponding virtual H&E-stained images at various depths from **A**.

## Discussion

We have demonstrated the effectiveness of combining APMD-FFOCT with a CycleGAN-based deep learning approach for performing 3D virtual H&E staining on unprocessed specimens, enabling rapid intraoperative diagnosis. This novel technique has been applied to human CNS and breast tumors, which are typically not suitable for conventional frozen section analysis but necessitate intraoperative diagnosis to inform surgical decision-making. We firstly address the instability of D-FFOCT images, which is caused by the random environmental vibrations. This issue is most prominent in structures with high reflectivity but low dynamics. By employing active phase modulation to eliminate the influence of the random environmental vibration, APMD-FFOCT images are well stabilized. This method also offers good contrast to static tissues, such as tumor-associated collagen fibers and calcified tissues, while maintaining high sensitivity to dynamic tissues like tumor cells. As the hue and brightness of dynamic images represent metabolic index of tissues, this adoption greatly improves the interpretability and reliability of the images. In addition, APMD-FFOCT also enhances the continuity and consistency in image stitching, which is important for accurate pathological diagnosis.

We then compare APMD-FFOCT images and H&E-stained images across various types of CNS tumors. Standard H&E staining is a fundamentally histopathological technique that highlights tissue structures and cellular details. Hematoxylin stains cell nuclei blue or purple, highlighting DNA and RNA, while Eosin colors the cytoplasm and extracellular matrix in shades of pink and red. This differentiation allows pathologists to examine tissue morphology, discern normal from abnormal structures, and assess cellular features crucial for diagnosing diseases. Correspondingly, APMD-FFOCT, a label-free imaging technique, renders nuclei the brightest due to their high metabolic dynamics, followed by the cytoplasm with a lesser brightness reflecting its weaker metabolic activity. The extracellular matrix, typically more metabolically inert, can still be visualized distinctly when enhanced through active phase modulation. Therefore, although arising from different contrast mechanisms, APMD-FFOCT provides pathological information comparable to standard H&E-stained images. Particularly for calcified structures, like the psammoma bodies, APMD-FFOCT offers images of superior quality compared to H&E staining, as it avoids the need for intricate tissue processing procedures.

While APMD-FFOCT provides quick acquisition of high-contrast pathological images, it also introduces an extra learning cost for pathologists, impeding swift diagnostic decisions. This complexity stems from the fact that pathologists are traditionally accustomed to interpreting diagnoses from conventional histological stains. To overcome these challenges, we have implemented a CycleGAN-based deep learning to perform virtual H&E staining on the APMD-FFOCT images. It is worth to note that the image stabilization achieved by APMD-FFOCT is crucial for virtual staining. Without it, the instability in traditional D-FFOCT might lead to the generation of erroneous information during virtual staining. The performance of our virtual H&E staining method has been validated by CNS tumor and IDC samples. The virtual H&E images, on one hand, faithfully reflect the details presented in the original APMD-FFOCT images, such as the morphology and distribution of cell nuclei, cytoplasm, and extracellular matrix. On the other hand, their staining style is highly consistent with that of real H&E staining. Moreover, generating each APMD-FFOCT image typically involves capturing 500 raw images within a 5-second window to mitigate speckle noise and ensure high image quality. Nonetheless, thanks to the robustness of our model, reducing the acquisition time and the number of raw images to 1 second and 100 frames still yields almost the same virtual H&E images. As the processing procedures, which include the generation of APMD-FFOCT images and their conversion into virtual H&E-stained images, can occur concurrently with the acquisition of raw images, the overall imaging duration required to produce virtual H&E-stained images of wide areas is substantially shortened.

Because nuclei of tumor cells exhibit significantly higher metabolic activities, their contrast is more pronounced in APMD-FFOCT images compared to other label-free imaging techniques (*31-34*). This advantage helps effectively mitigate issues of missing or erroneously generated nuclei in deep learning transformations, thereby enhancing the reliability of the resulting virtual H&E-stained images. Moreover, nuclear pleomorphism is a critical but challenging parameter to assess histological grade, exhibiting low reproducibility among even expert pathologists (*47*). This uncertainty is primarily attributed to the limitations of 2D histological specimen analysis (*48*), as current histology only provides typically the information of a certain section of the cell and lacks three-dimensional image of the natural living cell . Due to low-coherence gating, APMD-FFOCT is able to capture high axial resolution and label-free tomographic imaging without the need of slicing. Deep learning algorithm can then transform these into 3D virtual H&E-stained volumes and provide a clearer and more intuitive visualization of nuclear pleomorphism and subnuclear structures. Therefore, this new technique will surely enhance the assessment's accuracy and reliability effectively.

From a cost and size perspective, APMD-FFOCT requires only an LED as the light source, eliminating the need for expensive and bulky pulsed lasers. Moreover, its adoption of full-field imaging negates the necessity for a complex galvanometer scanning system, further simplifying the system's architecture. Importantly, this method does not require any labeling or fixation of tissues, and the light power irradiated onto biological tissues is only 1.2mW, ensuring no damage to the tissue and allowing for its reuse. Compare to traditional D-FFOCT, no additional time is required in APMD-FFOCT imaging. Further improvements and exploration should be made toward better clinical use. For example, APMD-FFOCT remains susceptible to low-frequency disturbances, like air flow or human activity. Future efforts could focus on reducing the impact of low-frequency vibrations by improving the instrument's anti-interference capabilities and implementing denoising algorithms. Moreover, the correlation between subnuclear structures observed in APMD-FFOCT and those in H&E-stained images requires additional investigation. Finally, the frame acquisition time, field of view size, and axial resolution of APMD-FFOCT images are independent factors. This independence enables further reduction in imaging time through the use of high-resolution objectives with a large field of view.

In summary, the development of APMD-FFOCT marks a significant advancement in stabilizing traditional D-FFOCT and enhancing the contrast of biological tissues with low metabolic activity. This lays the groundwork for utilizing deep learning to create 3D virtual H&E-stained images. Furthermore, the use of deep learning significantly reduces overall imaging time by decreasing the number of raw images required. We have also demonstrated the effectiveness of the combination of APMD-FFOCT and virtual H&E staining on human CNS and breast tumors for fast diagnosis in intraoperative histology. We believe this approach could greatly assist in pathological diagnosis and provide rapid feedback for intraoperative decision-making.

## Materials and Methods

### Experimental setup

Light from an LED (Thorlabs M565L3, with a central wavelength of 565 nm and a full-width at half maximum of 104 nm) is divided equally into the sample and reference arms via a 50/50 non-polarizing beam splitter (BS013, Thorlabs). This light is focused at the rear plane of the objectives to achieve Köhler illumination, which provides uniform lighting across the field of view. Identical high-magnification water immersion objectives (Nikon NIR APO 20x 0.5 NA) are employed to attain a lateral resolution of approximately 0.7 μm over a 500 × 350 μm$^2$ field of view. Using water as a medium allows for refractive index matching to reduce surface reflections from the objective lens and cover slides. Simultaneously, it minimizes the separation between the focal plane and the coherence plane when adjusting imaging depth. The axial resolution is ~1 μm, determined by the light source's coherence length.

In the sample arm, freshly excised tissues are placed on a custom sample holder, with a cover glass to flatten the tissue. The height of the cover glass can be finely adjusted using a threaded ring, which is then locked in place by a clasp to ensure stability. The holder is set on a five-dimensional control stage, with three for translational adjustments and two for angular corrections, ensuring the cover glass remains parallel to the focal plane for consistent imaging depth during image stitching. In the reference arm, a YAG (Yttrium Aluminum Garnet) reference mirror, modulated by an underlying PZT (TA0505D024W, Thorlabs), induces 25Hz phase modulation during imaging. This arm is attached to a high-precision electric translation stage (M-VP-25XL, Newport) to fine-tune the optical path difference between the arms, aligning the objective's focal plane with the coherence plane for optimal imaging.

Scattered light from subcellular particles and the reference mirror recombine at the beam splitter, and the merged beam is then focused onto a camera (MV4-D1600-S01-GT, Photon Focus) through a tube lens. The entire apparatus is stationed on an active vibration isolation platform (VCM-S400, Jiangxi Shengsheng) to mitigate environmental disturbances.

**Data acquisition, processing and image generation**

In our study, we employed custom-designed C++ software to collect all APMD-FFOCT datasets, facilitating three-dimensional scanning capabilities. The acquisition time of a dynamic image is set at 5 seconds (except Fig. 5), with a frame rate of 100 fps, to capture a wide range of intracellular movements. Each acquisition produces a tensor sized (550, 800, 500), where 550 × 800 represents the sensor pixels after pixel binning (2,2), and 500 denotes the count of frames captured. Binning reduces pixel count, conserving storage, and quadruples the quantum well depth, enhancing the SNR. Each pixel yields a time-domain intensity trace of 500 points across 5 seconds, which is then Fourier-transformed to derive the frequency spectrum. The spectrum's lowest and highest detectable motion frequencies are 0.2 Hz and 50 Hz, determined by the total acquisition duration and the camera's frame rate. To visualize the dynamic characteristics of metabolic activities, we utilized the Hue-Saturation-Value (HSV) color model. The hue, indicating the average fluctuation velocity, is calculated from the spectral centroid.

$$\text{Hue} = \frac{\sum_{n=1}^{N} f(n) \cdot S(f(n))}{\sum_{n=1}^{N} S(f(n))}$$

$N$ is the number of points in the spectrum, $f(n)$ is the value of frequency, representing each frequency point in discrete Fourier transform. $S(f(n))$ is the spectral intensity at the frequency $f(n)$. This formula calculates the weighted average of each frequency point in the spectrum.

The brightness of the image is determined by the integration of the frequency spectrum, excluding the zero-frequency portion (represents static background), which represents fluctuation amplitude.

$$Brightness = \sum_{n=1}^{N} S(f(n))$$

Saturation, inversely related to the standard deviation of the frequencies, is calculated next. A wider spectrum indicates lower saturation, which, to an extent, mirrors the complexity of motion patterns. All data processing is executed via our custom software. Saturation is quantified as follows, where $sc$ denotes the spectral centroid:

$$\text{Saturation} = 1/\sqrt{\frac{\sum_{n=1}^{N}(f(n)-sc)^2 \cdot S(f(n))}{\sum_{n=1}^{N}S(f(n))}}$$

Then, the above parameters are scaled by linear transformation. Finally, the HSV image is transformed in the RGB color space to display.

**Converting APMD-FFOCT images into H&E-stained Images with CycleGAN**

The raw APMD-FFOCT datasets were first processed using the above-mentioned Fourier-transform method. Frequency spectra at 25 Hz were isolated to enable flexible adjustment of hue and brightness for static tissues, such as collagen fibers and calcified tissues. These tissues were normalized to a moderate intensity. Furthermore, since the nuclei are brighter than the cytoplasm, the images were then converted to grayscale and underwent an inversion process to match standard H&E-stained images. Then, we used a trained deep neural network to convert gray APMD-FFOCT Images into H&E-stained Images.

We utilized CycleGAN, a type of Generative Adversarial Network (GAN) known for its effectiveness in image-to-image translation tasks without the need for paired images. The architecture of CycleGAN is composed by two generators ($G$: OCT → HE, $F$: HE→ OCT) (*30*) and two discriminators($D_{OCT}$ and $D_{HE}$). The purpose of virtual H&E staining is to determine the most suitable generator $G$: OCT → HE. The CycleGAN's success mainly contributed to its cycle-consistency loss, which ensures that an image can be translated from one domain to another and back again, with the goal of the returned image being indistinguishable from the original. So, the generators could generate the images that mimic the statistical properties of target domain. By using the cycle-consistency property of CycleGAN framework, the cycle-consistency loss function was defined as:

$$L_{cycle} = |F(G(OCT) - OCT)| + |G(F(HE) - HE)|$$

The least-squares adversarial losses were combined with cycle-consistency loss to train the generators, so the total loss function was defined as:

$$L_G = (1 - D_{HE}(G(OCT)))^2 + (1 - D_{OCT}(F(HE)))^2 + \lambda L_{cycle}$$

where $\lambda$ is the regularization parameter which was set to 10.

As opposed to the generators, the discriminators were used to distinguish between the real and the generated images, and the loss function for $D_{OCT}$ and $D_{HE}$ were defined as:

$$L_{D_{HE}} = D_{HE}(G(\text{OCT}))^2 + (1 - D_{HE}(\text{OCT}))^2$$

$$L_{D_{OCT}} = D_{OCT}(F(\text{HE}))^2 + (1 - D_{OCT}(\text{OCT}))^2$$

The detailed architecture of CycleGAN was descried by Zhu. et. al. The training dataset contains 3600 APMD-FFOCT images and 3450 H&E tiles of size 256 × 256 pixels, including human CNS tumor and breast tumor images patches. And 400 APMD-FFOCT images of size 800 × 550 pixels were used for testing. The training process was performed on NVIDIA GTX 3080 10GB GPU. The loss function was optimized using the Adam solver, with an initial learning rate of 0.0001 and batch size of 8. After 200 training epochs, the resulting generator G model's weights were used to convert APMD-FFOCT Images into H&E-stained Images.

**Sample preparation**

The study procedures of CNS tumors received approval from the Ethics Committee of Beijing Tsinghua Changgung Hospital, and the study procedures of Breast tumors were approved by the Ethics Committee of Peking University People's Hospital. Informed consent was obtained from each patient undergoing cancer surgery before imaging. Fresh tumor tissue was promptly collected from the tumor bed of the excised surgical specimen, ensuring the presence of at least one smooth and flat surface suitable for APMD-FFOCT imaging. The excised tissue was kept moist with saline and the container was stored on ice until imaging. All data were obtained within 2 hours after excision at room temperature. After APMD-FFOCT imaging, the tissue was fixed in formalin for paraffin H&E pathology. A pathologist with expertise in histology provided a pathological diagnosis for each tissue based on the corresponding H&E slide.

**Acknowledgments**

**Funding:**

National Natural Science Foundation of China with Grant No. 61575108, 61905015, 61975091 and 62275018

Bio-Brain+X' Advanced Imaging Instrument Development Seed Grant

Tsinghua Precision Medicine Foundation

**Author contributions:**

Z.C.Y. designed the system and software, collected and analyzed the data, and wrote the original manuscript; B.H. processed virtual H&E staining and edited the manuscript; Y.Z.Y. prepared CNS tumor samples; S.W.Z. prepared IDC tumor samples; P.Q.Y., Z.Y.C., and Z.W.H. assisted with data acquisition; Y.J.S., R.Z.X., and C.M.W. validated the system; S.W. explained the IDC tumor data and reviewed the manuscript; G.H.W. explained the CNS tumor data and reviewed the manuscript; P.X. supervised and directed the project, and reviewed and edited the manuscript.

**Competing interests:**

Authors declare that they have no competing interests.

**Data and materials availability:**

All data supporting the findings of this study are available within this paper. The raw OCT data, due to their large file size, are available from the corresponding author upon request. The original code for CycleGAN is available at https://github.com/junyanz/pytorch-CycleGAN-and-pix2pix. We applied this code to our dataset with the customized settings described in Methods.